# Scattering medium: randomly packed pinhole cameras

Honglin Liu[1*], Xin Wang[1], Puxiang Lai[2*], Zhentao Liu[1], Jianhong Shi[3] and Shensheng Han[1,4]

[1]Key Laboratory for Quantum Optics, Shanghai Institute of Optics and Fine Mechanics, Chinese Academy of Sciences, Shanghai 201800, China

[2]Department of Biomedical Engineering, Hong Kong Polytechnic University, Hong Kong SAR, China

[3]State Key Laboratory of Advanced Optical Communication Systems and Networks and Center of Quantum Sensing and Information Processing (QSIP), Shanghai Jiao Tong University, Shanghai, 200240, China

[4]Hangzhou Institute for Advanced study, University of Chinese Academy of Sciences, Hangzhou 310024, China

* Corresponding email: hlliu4@hotmail.com and puxiang.lai@polyu.edu.hk.

**Abstract**: When light travels through scattering media, speckles (spatially random distribution of fluctuated intensities) are formed due to the interference of light travelling along different optical paths, preventing the perception of structure, absolute location and dimension of a target within or on the other side of the medium. Currently, the prevailing techniques such as wavefront shaping, optical phase conjugation, scattering matrix measurement, and speckle autocorrelation imaging can only picture the target structure in the absence of prior information. Here we show that a scattering medium can be conceptualized as an assembly of randomly packed pinhole cameras, and the corresponding speckle pattern is a superposition of randomly shifted pinhole images. This provides a new perspective to bridge target, scattering medium, and speckle pattern, allowing one to localize and profile a target quantitatively from speckle patterns perceived from the other side of the scattering medium, which is impossible with all existing methods. The method also allows us to interpret some phenomena of diffusive light that are otherwise challenging to understand. For example, why the morphological appearance of speckle patterns changes with the target, why information is difficult to be extracted from thick scattering media, and what determines the capability of seeing through scattering media. In summary, the concept, whilst in its infancy, opens a new door to unveiling scattering media and information extraction from scattering media in real time.

Seeing clearly through scattering media (e.g. biological tissues) has been desired for long yet is always a big challenge. The usual practices are either to adopt less scattered radiation or mechanical waves, or to tag and compensate for distorted wavefronts to achieve turbidity suppression. Examples of the former category include, but are not limited to, X-ray imaging, computed tomography (CT), single-photon emission computed tomography (SPECT), ultrasound imaging, photoacoustic imaging [1], etc. For the latter category, researchers have recently proposed approaches like wavefront shaping [2], optical phase conjugation [3-7], scattering matrix measurement [8,9], and so on.

Different from the aforementioned approaches, speckle autocorrelation imaging extracts object information directly from speckle patterns under incoherent illumination [10]. The underlying principle is that speckle pattern results from the convolution of the object and the speckle point spread function (sPSF), the speckle under a point illumination; thus, autocorrelation of speckle patterns equals to the autocorrelation of the object itself. Since its first report in 2012 [10], speckle autocorrelation imaging has attracted intense explorations. In this method, an image of a target can be retrieved from speckle patterns through algorithms, such as Gerchberg–Saxton and HIO alternative algorithms. Visually, the speckle pattern itself is randomized. However, under scrutinization the profiles of the hidden object can be seen in the speckle pattern, as shown in Fig.1, like small boats floating amongst massive surges. Similar observations can be easily found in papers regarding speckle autocorrelation imaging [11-18]. Moreover, if the object profile disappears from the speckle pattern, with little chance the image can be retrieved in subsequence. Another well accepted concept in the field is that a scattering medium is a scattering lens [19,20]. But different from conventional lenses, a scattering lens has both infinite depth-of-field and depth-of-focus. What is the principle behind these phenomena?

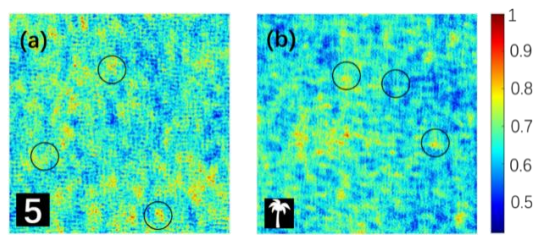

Fig. 1 Examples of speckle patterns corresponding to different objects (See Fig. S1 in Supplementary Information (SI) for its grey counterpart for a different visual preference). The insets on the bottom left corners of (a) and (b) are the object photographs, respectively. In both figures, the intensity distributions are randomized, but their appearances are apparently different. Black circles highlight the profiles that are likely linked with the target images, albeit blurred and spatially reversed.

Aiming to understand these phenomena, inspirations from pinhole camera are conceived. Pinhole camera, earliest mentioned in Mohist (Mozi, China, ~500 BC), is one of the oldest image-making technologies, and it is absolutely distinctive from correlation imaging in terms of principle and application scenario. Interesting connection, however, exists between these two schemes: a scattering medium can be considered as numerous randomly packed pinhole cameras; consequentially, a speckle pattern can be considered as a superposition of many pinhole images with randomly shifted centers. To rationalize the statement, in this paper, randomly packed pinhole cameras are first analyzed on a plane and then extended to 3 dimensions. The maximum image angle of a volumetric scattering medium is derived, yielding a result coincident with the range of the angular memory effect [21-23]. A model is developed to investigate the evolution of speckles with increased number of pinhole cameras, demonstrating that a scattering medium in fact functions as randomly packed pinhole cameras.

Considering that tremendous light is either blocked or absorbed in a pinhole camera scheme, it is different from an actual turbid medium where most photons are multiply scattered and finally escape in all directions. To further scrutinize the idea, a phase pinhole camera model is built to simulate speckle autocorrelation imaging of a phasemask; the hypothesis is that a scattering medium can be mimicked by one or multiple phasemasks [24-26]. By exploring the geometry relationship between the target, the scattering medium, and the speckle pattern, a solution is found to localize and profile a target quantitatively through a scattering medium in real time, which is impossible with all existing methods.

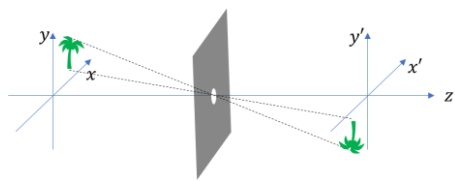

Fig. 2 Illustration of a pinhole camera system with equal distances for object and image

**Principles**

First, let's consider a pinhole camera system with equal distances for object and image, as shown in Fig.2, for simplicity. Thus, a reversed real image of the same size as the object can be obtained on the imaging plane, with an intensity of

$$I(x',y') = \iint O(x,y)\delta(x+x', y+y')dxdy = O(-x',-y'), \quad (1)$$

where $O(x,y)$ is the intensity distribution of the object under incoherent illumination, and $\delta(x+x', y+y')$ is the ideal response function of the pinhole camera. Eq.(1) has another form,

$$I(x',y') = [O\star\delta](x',y'), \quad (2)$$

where $\delta(x',y')$ is the point spread function (PSF) and $\star$ is an operator of correlation. Note that there is no reversion to the PSF; the intensity on the image plane is a correlation, instead of convolution, of the object and the PSF (for more details, please refer to Section B in SI). Under geometric assumption, the diameter of the pinhole is zero. If the center of the pinhole is shifted to $(x_p, y_p)$, the corresponding PSF is $\delta(x' - 2x_p, y' - 2y_p)$, and Eq.(2) turns to be

$$I(x',y') = O(x',y')\star\delta(x' - 2x_p, y' - 2y_p). \quad (3)$$

The autocorrelation of the pinhole image is thus

$$\begin{aligned} I(x',y')\star I(x',y') &= [O(x',y')\star\delta(x'-2x_p, y'-2y_p)]\star[O(x',y')\star\delta(x'-2x_p, y'-2y_p)] \\ &= [O(x',y') \star O(x',y')]\star[\delta(x'-2x_p, y'-2y_p)\star\delta(x'-2x_p, y'-2y_p)] \\ &= [O(x',y') \star O(x',y')]\star\delta(x',y') \\ &= O(x',y') \star O(x',y'). \end{aligned} \quad (4)$$

Obviously, no matter where the center is, the autocorrelation of each pinhole image is the same, and can be added in phase. Next, let's consider a black screen with many randomly distributed pinhole cameras. The intensity on the image plane is

$$\begin{aligned} I(x',y') &= \sum_i O(x',y')\star\delta(x'-2x_{pi}, y'-2y_{pi}) \\ &= O(x',y')\star\sum_i \delta(x'-2x_{pi}, y'-2y_{pi}), \end{aligned} \quad (5)$$

where $(x_{pi}, y_{pi}), i = integers$ is the center of each pinhole camera. Because of the zero width of a Delta function, the autocorrelation of $\sum_i \delta(x'-2x_{pi}, y'-2y_{pi})$ equals to $\delta(x',y')$ and the intensity autocorrelation is still the autocorrelation of the object. As the number density of pinhole cameras increases, the intensity on the image plane evolves from sparsely distributed images into a speckle pattern.

Now, let's turn to speckle autocorrelation imaging, where speckle pattern on the image plane can be denoted by [10-18]

$$I(x',y') = \sum_i O(x',y')\star sPSF, \quad (6)$$

where $sPSF$ denotes the speckle point spread function of the scattering medium. Intuitively, the $PSF$ of randomly aligned pinhole cameras overall is a counterpart of the $sPSF$. In other words

$$sPSF = \sum_i \delta(x'-2x_{pi}, y'-2y_{pi}). \quad (7)$$

Therefore, the scattering medium can be treated as an assembly of randomly packed pinhole cameras, and the corresponding speckle pattern as a superposition of numerous randomly shifted images.

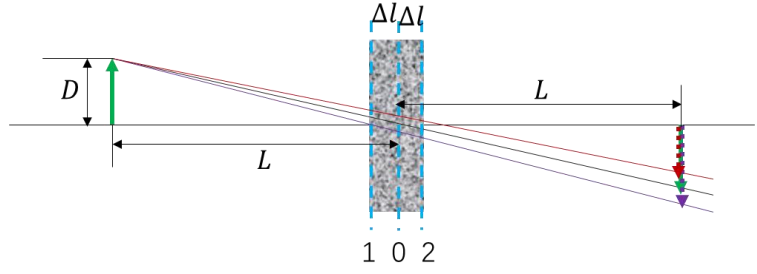

Fig. 3 Configuration of a scattering medium as 3D randomly packed pinhole cameras. An object is imaged by pinhole cameras onto different planes. $2\Delta l$ is the thickness of the medium and $L$ is the distance between the object and the center of the medium. The distance between the central plane of the medium and the CCD plane is also $L$. The height of the object is $D$.

Table 1 Magnification and image height on different planes.

|  | Magnification | Image height |
|---|---|---|
| Phase mask 0 | $M = 1$ | $D$ |
| Phase mask 1 | $M = \frac{L+\Delta l}{L-\Delta l}$ | $\frac{L+\Delta l}{L-\Delta l} D$ |
| Phase mask 2 | $M = \frac{L-\Delta l}{L+\Delta l}$ | $\frac{L-\Delta l}{L+\Delta l} D$ |

Note that the above derivations have dealt with pinhole cameras on a plane, which, however, sees limitations as the thicknesses of most scattering media are not negligible. A more reasonable model should consider cameras randomly distributed in a 3D space. Practically, we can slice the space into many layers that are sufficiently thin to be eligible as 2-D planes, with each containing many pinhole cameras, as shown in Fig. 3. For each plane, a speckle pattern is produced on its respective image plane. Thus, the magnifications of pinhole images are plane dependent. In Table 1, the magnifications and image height of three representative planes are listed. As the final speckle pattern is a supposition of speckle patterns generated by each plane independently, a series of images can be reconstructed from collected speckle patterns. But only when the difference of image sizes does not exceed half of a wavelength, the images are indistinguishable. That is,

$$\frac{L+\Delta l}{L-\Delta l} D - \frac{L-\Delta l}{L+\Delta l} D \le \frac{\lambda}{2}, \tag{8}$$

when the speckle autocorrelations of different planes can be added in phase. Considering $\Delta l \ll L$, we can have

$$\frac{D}{L} \le \frac{\lambda}{8\Delta l}, \tag{9}$$

where $\theta = D/L$ is the angle of object visualization on the medium, being limited within $\frac{\lambda}{8\Delta l}$, which is a coincidence with the angular memory effect range $\theta = \frac{\lambda}{2\pi\Delta l}$ [21,22].

**Results**

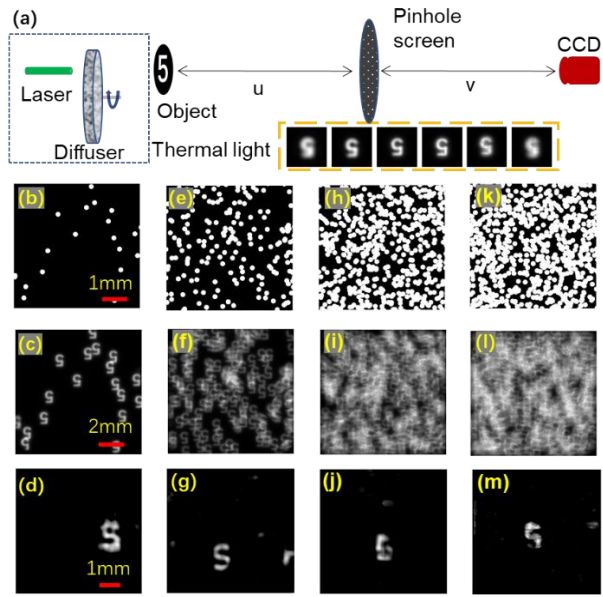

Fig. 4 Simulation of pinhole camera imaging through amplitude pinhole screens. (a) Configuration of the system. $u = 6\ cm$, $v = 6\ cm$, the height of the digital object is 1 mm. In the orange dash box shows the pinhole images with different pinhole radii of 50, 70, 90, 110, 120, and 140 $\mu m$, successively. Obviously, optimum image is obtained around 100 $\mu m$ radius. (b-m) are the simulated distributions of the pinholes on the screen (second row), light pattern recorded on the CCD plane (third row), and reconstructed images (fourth row), respectively, when the number of pinholes gradually increases from column to column (N=20, 200, 600, and 800).

To demonstrate the derivations above, we implement a pinhole camera system (Fig. 4) to simulate the detected light patterns and reconstructed images when the radius and number of pinholes are gradually changed. A 532 nm laser impinges on a rotating ground glass diffuser to generate incoherent light, which transmits through a digit number "5". On the pinhole screen there are a series of randomly distributed pinholes with a radius of 100 $\mu m$. The second row are simulated distributions of the pinholes on the screen, with 20, 200, 600 and 800 pinholes randomly distributed within a field of view of 5×5 mm$^2$, respectively. The third row are the simulated light patterns recorded on the CCD plane, and the fourth row are reconstructed images. As seen, with sparse pinholes (e.g., N=20), the images on the CCD plane are clearly distinguishable (Fig. 4c), and the digit image can be reconstructed. When the number of pinholes increases, more and more images are superposed, resulting seemingly random speckle patterns, where the images are no longer discrete (Figs. 4f, i, l). But with retrieval algorithms [11, 27-30] (see Section C in SI for more details), the digit can still be reconstructed (Figs. 4g, j, m), albeit with increased noise possibly caused by the truncations of four edges of the speckle patterns and the cross talk among overlapped pinholes.

Considering most turbid media of interest (e.g., biological tissue) are scattering dominant, simulations are also implemented based on one or several phase screens [25,26]. A model of phase pinhole camera is designed by positioning one or multiple small phase disks on a transparent plate; one phase disk corresponds to one pinhole, and its phase can be different values, as shown in the first row of Fig. 5. As seen, a reversed image of the object can be obtained, as long as the phase value of the disk is not the same or $2n\pi$ ($n$ is an integer) different with that of the background. Notably, the background of light pattern reaching the phase pinhole camera is bright (the second row of Fig. 5), which is different from amplitude pinhole cameras. As the number density of pinhole increases, the amplitude pinhole screen converges to a clear aperture, while phase pinhole screen has no such a problem. An example of phasemask with $8 \times 10^3$ phase disks randomly located on a flat phase plate is shown in Fig. 5f, which leads to a random speckle pattern on the CCD plane as illustrated in Fig. 5g (see Section D in SI for more details about generated phasemasks). Nevertheless, an image of the digit object can still be extracted (Fig. 5h).

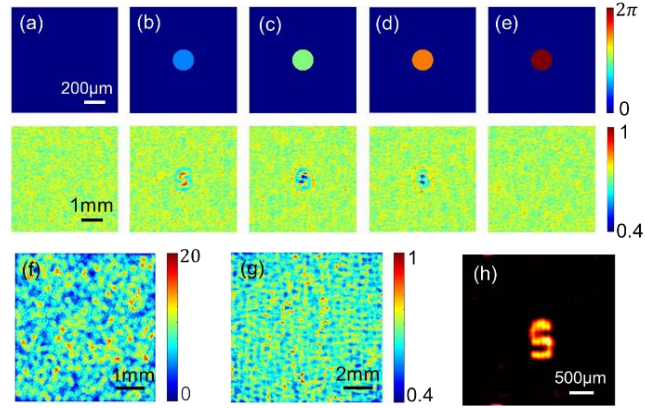

Fig. 5 Simulation results of randomly packed phase pinhole cameras. (a-e) The upper row illustrates phase screens with one circular bump of uniform phase (0, $\pi/2$, $\pi$, $3\pi/2$, and $2\pi$, respectively) at the center; the lower row are the corresponding intensity patterns on the detection plane. (f) An example of phase screen with $8 \times 10^3$ randomly distributed phase bumps within a field of view of 5×5 mm². The diameter of each disk is 200 $\mu m$, and the phase step height is $\pi/4$. (g) is the corresponding light pattern on the CCD plane after normalization and (h) is the reconstructed digit image.

All these confirm once again that a scattering medium can function as an assembly of randomly packed pinhole cameras, and speckle pattern is a superposition of countless randomly shifted pinhole images. Thus, it is natural to see profiles of hidden objects from speckle patterns, whose differences in visual appearance are new evidence that elementary cells matter in determining the macroscopic world. Also, as pinhole cameras have infinite depth-of-field and depth-of-focus, it intuitively supports similar claims for scattering lens [19,20]. However, for thick scattering media (containing many $\Delta l$ in Fig. 3), the sizes of images in response to pinholes at different layers of the medium might be too diverse to spot the object profiles in the speckle pattern. According to Eqs. (8-9), once the object exceeds the maximum imaging angle, *a.k.a.* the memory effect range, there is little chance to find a profile or retrieve the object through algorithms.

The findings above also inspire the wisdom to treat speckle pattern as a projection of the target through the scattering aperture, and their geometry relationship can be used to quantify and localize the target, as shown in Fig. 6 through both simulation and experiment. A homemade digit "5" with a height of 320 $\mu m$ is selected as the target and a ground glass diffuser (Thorlabs DG20-600-MD) is used as the scattering medium. The diameter of the transmission window $D_s$ =2.0 mm. Other settings are the same as those in Fig. 4a. In Fig. 6a, the dark green denotes the projection of the target (the size is neglected under paraxial approximation) through the diffuser, and the light green represents the diffraction. Fig. 6b is the simulated speckle pattern with a centrosymmetric profile. The center can be calculated by weighted centroid localization algorithm [31] to be (3.3, 0.8) $\mu m$, being very slightly different from its actual location at (0, 0) $\mu m$. Not surprisingly, the projection area has a higher intensity than the background does. By selecting proper circles in Fig. 6c, we can estimate the diameter of the projection area $D_p$. The magnification $M = D_p/D_s - 1$. Although selection of the circle maybe subjective and inconsistent, the deviation rate of $M$ can be controlled within 5% after training, as reflected in Fig. 6c. Since the image distance $v$ (with respect to the diffuser) is measurable, the position of the object $u$ can be calculated. After image reconstruction, we can furthery calculate the size of the target. Such a capability enables one to monitor and trace moving targets through scattering media in real time without any invasion, perturbation, or prior information, which has not yet been reported.

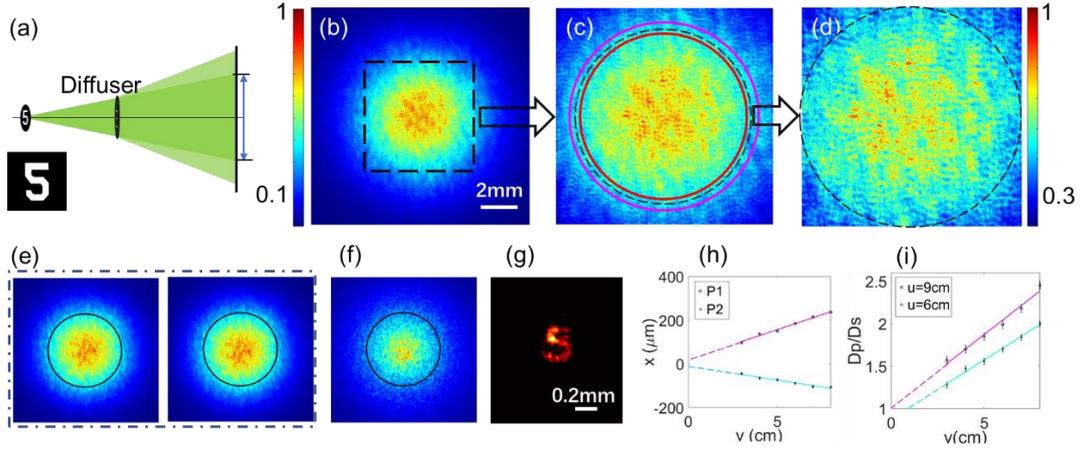

Fig. 6 Quantitative picturing and localization of a target through a scattering medium. (a) Geometric illustration of target projection and light diffraction through the medium (a diffuser in experiment). The insert is the photograph of the digit object. (b) The simulated speckle pattern on the imaging plane. (c) The dash square in (b) is zoomed in. The black dash circle denotes the real projection area; the inner and outer circles are estimations with -5% and +5% deviation rates, respectively. (d) Further zoom in of the black dash circle in (c). (e) Simulated speckle pattern before (on the left) and after (on the right) the object shifted along y axis by 150.0 $\mu m$. No apparent variation in speckle appearance can be seen, but the calculated center at (-3.0, 154.5) $\mu m$ indicates the target movement along y axis. (f) Speckle pattern detected in experiment with settings similar to those in (b). (g) The reconstructed target image from the projection rather than diffraction area of (f). (h-i) In experiment, the digit target is first positioned at P1 =80.0 $\mu m$, then moved along x axis to P2=-180.0 $\mu m$. For both object positions, speckle patterns at different image distances v are recorded. The centers of these speckle patterns are plotted and linearly fitted as a function of v in (h). The slope of fitted curve is proportional to the position angle seeing from the diffuser, and ideally the two fitted curves in response to the two target positions cross at v=0. In (i), the ratio $D_p/D_s$ as a function of imaging distance $v$ at different object distances $u$ is shown. The inverse of the slope of the fitted curve represents the value of $u$. The calculated object distances for the magenta (u=6.0cm) and cyan (u=9.0cm) fitted linear curves are 5.8cm and 8.6cm, respectively.

**Discussion**

Thus far, the potentials of the new concept have been largely demonstrated. Three more aspects, however, need to be discussed. First, in derivations above, a pinhole is an ideal point, and there are no overlaps between adjacent points. Thus, the corresponding images are independent to each other, the superposition on the image plane is an algebraic addition, and each speckle pattern corresponds to a unique pinhole screen. In practice, however, pinholes have non-zero diameters; when the center-to-center distance between two pinholes is smaller than the diameter, overlaps occur. Such cross talks bring noise, deteriorating the quality of reconstructed images. Moreover, massive overlaps of phase pinhole bumps eventually lead to plateau of phase. The height of the biggest plateau represents the mean of the phasemask. Usually there are chances to find discrete phase pinhole bumps on different phase plateaus of a phasemask, which explains the emergence of the object profile out of the speckle background.

Second, the model has thus far focused on single scattering, where light interacts with only one pinhole plane. For thick media composed of multiple phasemask layers, light has more chances to interact with many pinholes successively. For single scattering, the input of all pinhole cameras is the object, while for multiple scattering, the input of a latter pinhole is the output of all pinholes prior to it. In this case, the autocorrelation of the output is complicate, encoded with the pinhole distribution information, but no longer the autocorrelation of the object alone. Therefore, the output of multiple scattering mainly contributes to the background noise in speckle autocorrelation imaging. As pointed out in caption of Fig.6 that the image can be reconstructed only from the speckle pattern in the projection area, which is an evidence for the model of information transmission through scattering media.

Third, it has been shown that a speckle pattern can be viewed as a superposition of many randomly shifted image cells. With different objects, the corresponding speckle patterns have different appearances because of the different elementary cells. Recently, deep learning has been standing out in imaging through scattering media [32-35]. However, the design of the network depends on experience and the training needs massive data. More importantly, how the encoded optical information is extracted remains a black box. The pinhole camera assumption shines a light to the underlying mechanics of deep learning. Although hidden, the object profile is preserved in the speckle pattern; the network is trained to identify the profile from random intensity fluctuations. For the massive training data, the pinhole centers of different speckle patterns follow the same statistical distribution, and the characteristics of a speckle pattern is its elementary cell rather than the statistical distribution. As sample thickness increases, the weight of single scattering light and hence the weight of characteristic information decrease, which explains the failure of deep learning for imaging through thick scattering media even with sufficient light flux. Therefore, with the pinhole camera assumption, we can better understand speckles from a new perspective, with which we may further improve the way to extract speckle characteristic information and design an efficient physical model rather than data driven network for deep learning. This could be a solution for the poor generalization capability of current deep learning.

In this work, a new concept is proposed and demonstrated to treat scattering media as assemblies of randomly packed pinhole cameras and speckle pattern as a superposition of numerous pinhole images with randomly shifted centers. The model may refresh or deepen our understanding of scattering media and speckle patterns from a new perspective. It also provides us a powerful tool to investigate how information transmits in scattering media and what determines the capability of speckle autocorrelation imaging and deep learning. The famous memory effect range is the maximum imaging angle of a volumetric scattering medium under the condition that the difference of image size does not exceed half of a wavelength. Lastly but not the least, the geometric relationship between the target, scattering medium, and speckle pattern allows one to quantify the location and size of targets hidden behind or within scattering media in single shot with no need for prior information, which is a unique capability that has never been obtained. In summary, the concept unveils a new pathway to understanding scattering media and scattered information deciphering. It may find immediate and future benefits to a variety of applications in complex environment, such as quantitative biological fluorescence imaging [36-40], precise treatment planning, navigation in fog, and so on.

**Acknowledgements**
H. Liu thanks Dr. Lihong V. Wang for inspiring discussion. This work was supported by the National Key Research and Development Program of China (No. 2016YFC0100602); National Natural Science Foundation of China (NSFC) (81930048,81671726, 81627805); Hong Kong Research Grant Council (25204416); Guangdong Science and Technology Commission (2019A1515011374).

**Author Contributions**
H. L. conceived the idea, did the simulation, designed the experiment, analysed the data, X. W. performed experiment and image reconstruction. P. L. optimized the model. H. L and P. L. wrote the manuscript. All revised the manuscript.

**Competing Interests**
The authors declare no conflicts of interest.

**Correspondence and requests for materials** should be addressed to H. L or P. L.

**Supplementary Information**

**A. A grey counterpart of Fig.1**

Fig.S1 is Fig.1 in grey format.

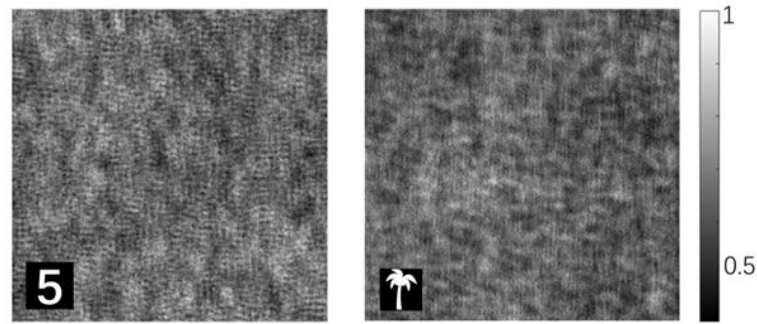

Fig.S1 A counterpart of Fig.1 in grey scale.

**B. Convolution or correlation?**

For an imaging system, such as a lens, the system response under incoherent illumination can be described by a point spread function ($PSF$). If an object O is inserted in the object plane, on the image plane we get an image $I = O \otimes PSF$, where $\otimes$ represents a convolution operator. We take it as the ground truth. Does it also apply for the speckle patten of an object through a scattering medium? Let's recall the definitions of convolution and correlation:

$$[O \otimes PSF](x') = \int_0^{x'} O(x) PSF(x' - x) dx, \quad (S1)$$

$$[O \star PSF](x') = \int_0^{x'} O(x) PSF(x + x') dx. \quad (S2)$$

Apparently, the main difference is that convolution has a reverse operation of the $PSF$, while correlation has no such operation. For a thin lens imaging system shown in Fig.S2, if the input is a point $\delta(x,y)$, the ideal system response is $\delta(x',y')$. In other words, $PSF(x',y') = \delta(x',y')$. For an object $O(x,y)$, if convolution applies, the image $I(x',y') = \iint_0^{x,y} O(x,y)\delta(x'-x, y'-y)dxdy = O(x',y')$; if correlation applies, the image $I(x',y') = \iint_0^{x,y} O(x,y)\delta(x+x', y+y')dxdy = O(-x',-y')$. The real image is then $O(-x',-y')$. So the real operation is correlation instead of convolution. Previously this problem was fixed by defining reduced coordinates in the object space with the magnification to be either negative or positive, according to whether the image is inverted or not [1].

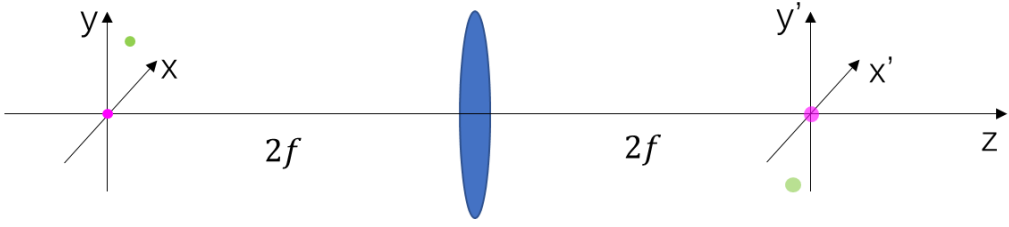

Fig.S2 Schematic illustration of a thin lens imaging system. $f$ is the focal length.

In Fig.S2 a slightly bigger magenta spot in x'-y' plane is the image of the magenta point in x-y plane. The magenta spot represents the $PSF$ of the system. For the green point in x-y plane, the system response is the green spot in x'-y' plane, same as the magenta spot except for a shifted center. For an object combined by many points, the image is the superposition of many corresponding spots on x'-y' plane. There is no 180° rotation (flip in 1D) of the $PSF$. For lens systems, usually the $PSF$ is symmetric, the convolution operation is equivalent to the correlation. For speckle autocorrelation imaging, the lens in Fig.S2 is replaced by a scattering layer, the $PSF$ turns to be a speckle pattern, which is not symmetric, within the memory effect range, the final intensity distribution is a superposition of many shifted speckle patterns, again with no rotation of the $PSF$. Some may argue that an image can be acquired by deconvolution even when the $PSF$ is a speckle pattern. The reason is as follows: $[\breve{O} \otimes PSF](x') = \int_0^x O(-x,-y)PSF(x'-x, y'-y)dx\,dy$, $\breve{O}$ is a 180° rotated $O$, $[\breve{O} \otimes PSF](x') = \int_0^x O(x,y)PSF(x+x', y+y')dx\,dy$, after deconvolution, $\breve{O}$ is obtained, and the rotation of $\breve{O}$ is usually done inadvertently. However, convolution and correlation are different operation essentially. Therefore, when a scattering medium is considered as an assembly of randomly packed pinhole cameras, the speckle pattern is a superposition of randomly shifted pinhole images. Pinhole images are inverted, a negative magnification should be introduced to use the convolution operation.

C.  **G-S algorithm for image retrieval**

For speckle autocorrelation imaging, we have

$$I \star I = O \star O. \tag{S3}$$

According to Weiner-Khinchin theorem, the spatial power spectral density (sPSD) of the object

$$S(k_x, k_y) = |\tilde{F}(O \star O)|, \tag{S4}$$

where $\tilde{F}$ is an operator of Fourier transform. Substituting Eq.(S3) into Eq.(S4) we get

$$S(k_x, k_y) = |\tilde{F}(I \star I)| = |\tilde{F}(I)|^2. \tag{S5}$$

From the speckle pattern, we can calculate the object's sPSD, *i.e.*, the amplitude of the object's Fourier transform. However, phase information is lost. A modified G-S algorithm with imbedded NNP (number of nonzero pixel), HIO (hybrid input-output) and ER (error reduction) algorithms is applied to retrieve the image. Fig.S3 shows the flow chart of the G-S algorithm.

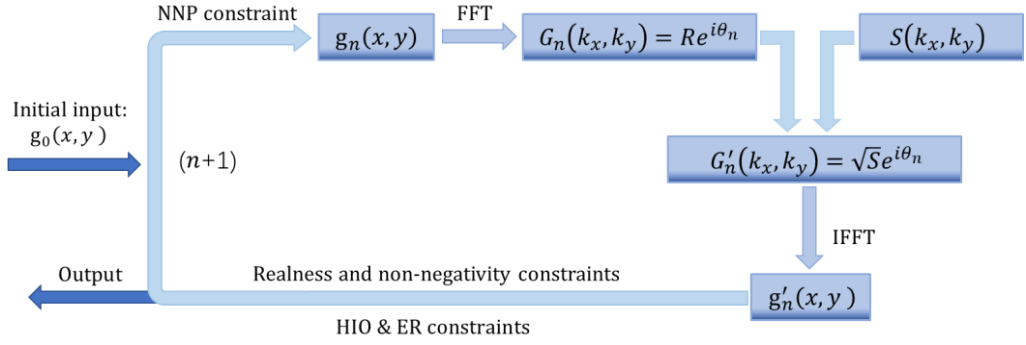

Fig.S3 Flow chart of the G-S iterative phase retrieval algorithm. The G-S algorithm starts with an initial random guess $g_0(x,y)$. It updates and saves the phase during the back-and-forth transformation of the real-domain and Fourier-domain, and follows three constraints: first, NNP constraint; second, HIO and ER constraints; third, to be constrained by $S$ in Fourier-domain, that is, the amplitude envelope is always constrained at $\sqrt{S}$. After a certain number of iterations, the recovered image is finally output.

The initial input $g_0(x,y)$ is a random 2D image in real space, $g_n(x,y)$ is obtained (for the first loop of iteration, n=1) after the NNP algorithm, *i.e.*, a kind of support constraint. $G_n(k_x,k_y)$ is the Fourier transform of $g_n(x,y)$ in spatial frequency space. Replacing its amplitude $R$ by $\sqrt{S}$, we obtain $G'_n(k_x,k_y)$. $\sqrt{S}$ is used as the amplitude constraint in frequency space. After inverse Fourier transform, we obtained $g'_n(x,y)$. For practical images in real space, all pixels have non-negative values, which can be used as amplitude constraints. HIO and ER algorithms are implemented to optimize the amplitude constraints to get $g_{n+1}(x,y)$ from $g_n(x,y)$ in each loop.

D. **What determines a phasemask?**

In the creation of a random phasemask, numerous phase bumps are randomly distributed within a predefined area. How the number, diameter, and step height of phase bumps affect the final distribution of the phasemask? We did simulations to show the evolution of phasemask with the number of phase disks at different diameters, 200 $\mu m$ in Fig. S4 and 40 $\mu m$ in Fig. S5. In both simulations the step height was set as $\pi/4$. Obviously, as the number $n$ increases, while the mean value of phasemask, which can be quantified by the center of the histogram, has a linear increment, the fluctuation range denoted by its width is proportional to the square root. Comparing to the 200 $\mu m$ diameter, the phasemasks generated by phase bumps with the 40 $\mu m$ diameter have finer grains. Since the product of phase disk area and the number $n$ are equal, the means of Fig. S4A and Fig. S5D have equal values, although the fluctuation of the latter is larger for finer resolution. Thus, in order to have a more continuous distribution in histogram, we can select one or several smaller step heights for the phase bumps.

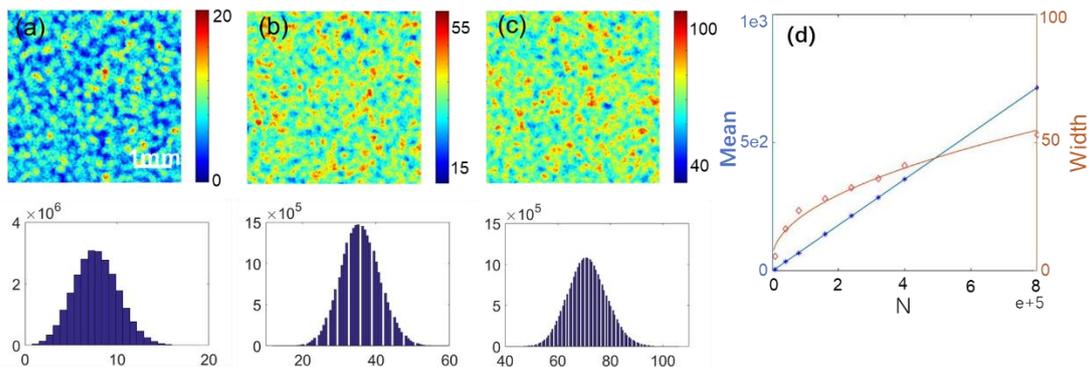

Fig.S4 Evolution of the phasemask with the number $n$ of phase bumps at the diameter of $200\mu m$. Within the $5\times5\ mm^2$ field of view, $8\times10^3$, $4\times10^4$ and $8\times10^4$ phase bumps are randomly distributed in the upper of (a-c), respectively. The lower shows the phase histogram of each phasemask correspondingly, where the bin width of histogram is always $\pi/4$. As $n$ increases, both the center and width of the histogram increase. (d) shows the trends of the center (corresponding to the mean) and width of generated phase mask with $n$.

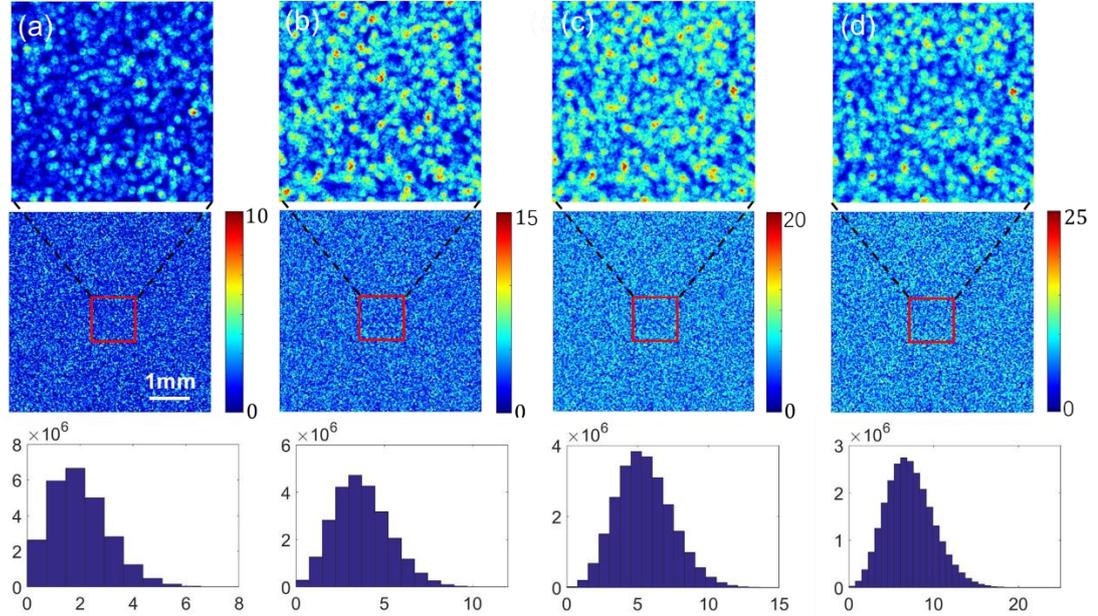

Fig.S5 Evolution of the phasemask with $n$ phase bumps at the diameter of $40\mu m$. Within the $5\times5\ mm^2$ field of view, $5\times10^4$, $1\times10^5$, $1.5\times10^5$ and $2.0\times10^5$ phase bumps are randomly distributed in the middle of (a-d), respectively. The upper is the zoom in of the red square. The lower shows the phase histogram of each phasemask correspondingly, where the bin width is also $\pi/4$. The histogram is non symmetric before the phasemask is fully grown.

Step phase bumps are the elementary cells of the above phasemasks, and elementary cells matter in determining their appearances, which are also speckles in terms of phase distribution. So, they look different from a phasemask based on a conventional model [2-4]. In the conventional model, the phasemask is generated from a Gaussian autocorrelation function $G(x,y) = \sigma^2\exp[-(x^2+y^2)/\kappa^2]$, where $\sigma$ is the height standard derivation and $\kappa$ is the transverse correlation length, of a typical ground glass model. To solve this problem, instead of the step phase bump, a Gaussian phase bump is used as the elementary cell. In fact, each grain of a ground glass disk can be considered as a Gaussian phase bump. Numerous grains are randomly located, corresponding to Gaussian phase bumps randomly distributed and superposed within an area. Fig.S6 shows the evolvement of phasemask with the number of phase bumps, with a conventional phasemask as a reference. Once the speckle pattern grows mature, the statistical distribution become stable while the fluctuation amplitude can be controlled by the amplitude and number of the phase bump cell.

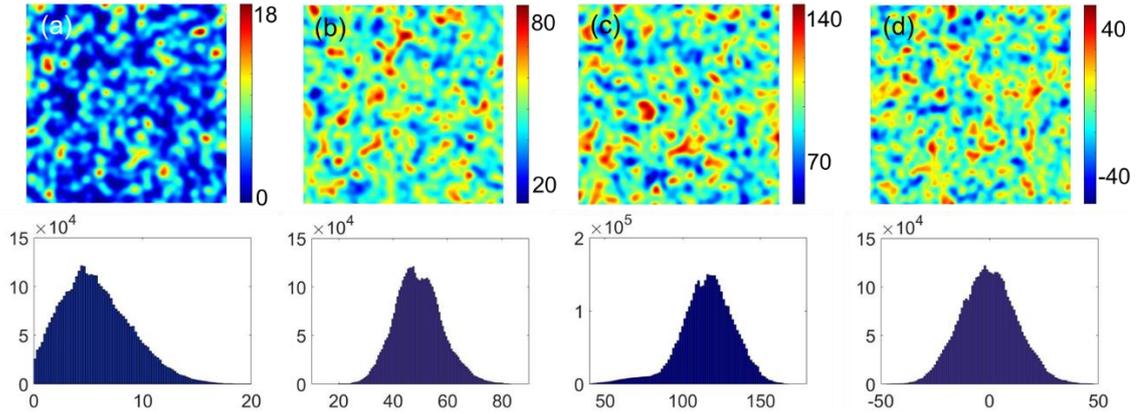

Fig.S6 Evolution of the phasemask with the number of Gaussian phase bumps with a height of $\pi$ and a radius $\kappa = 60\mu m$. Within a $2 \times 2\ mm^2$ square area, $10^3$, $1 \times 10^4$ and $2 \times 10^4$ Gaussian phase bumps are randomly distributed in the upper of (a-c), respectively. (d) is the a conventional phasemask generated under the constraint $G(x, y) = \sigma^2 \exp[-(x^2 + y^2)/\kappa^2]$ at $\sigma = 10\mu m$ and $\kappa = 60\mu m$. As the phasemask grows mature with a nearly symmetric histogram, it has the same appearance and statistical distribution as the conventional one except the amplitude. By changing the height and/or number density of the phase bump cell, the amplitude of generated phasemask can be altered.